\documentclass[sn-mathphys]{sn-jnl}% Math and Physical Sciences Reference Style
\jyear{2021}%

\theoremstyle{thmstyleone}%
\theoremstyle{thmstyletwo}%

\theoremstyle{thmstylethree}%

\newtheorem{thm}{Theorem}[section]

\newtheorem{defn}{Definition}[section]

\newcommand{\Pm}{{\rm Pm}}
\renewcommand{\Re}{{\rm Re\,}}
\newcommand{\Rm}{{\rm Rm\,}}
\newcommand{\Ha}{{\rm Ha}}
\newcommand{\be}{\begin{equation}}
\newcommand{\ee}{\end{equation}}	

\begin{document}

\title[Monotone energy stability of magnetohydrodynamics ...]{Monotone energy stability of magnetohydrodynamics Couette and Hartmann flows}

%%=============================================================%%
%% Prefix	-> \pfx{Dr}
%% GivenName	-> \fnm{Joergen W.}
%% Particle	-> \spfx{van der} -> surname prefix
%% FamilyName	-> \sur{Ploeg}
%% Suffix	-> \sfx{IV}
%% NatureName	-> \tanm{Poet Laureate} -> Title after name
%% Degrees	-> \dgr{MSc, PhD}
%% \author*[1,2]{\pfx{Dr} \fnm{Joergen W.} \spfx{van der} \sur{Ploeg} \sfx{IV} \tanm{Poet Laureate} 
%%                 \dgr{MSc, PhD}}\email{iauthor@gmail.com}
%%=============================================================%%

\author{\fnm{Giuseppe} \sur{Mulone}}\email{giuseppe.mulone@unict.it}

\affil{\orgdiv{Department of Mathematics and Computer Science (retired)}, \orgname{Università di Catania}, \orgaddress{\street{Viale Andrea Doria 6}, \city{Catania}, \postcode{95125}, \state{Italy}}}

%\affil[2]{\orgdiv{Department}, \orgname{Organization}, \orgaddress{\street{Street}, \city{City}, \postcode{10587}, \state{State}, \country{Country}}}
%
%\affil[3]{\orgdiv{Department}, \orgname{Organization}, \orgaddress{\street{Street}, \city{City}, \postcode{610101}, \state{State}, \country{Country}}}

%%==================================%%
%% sample for unstructured abstract %%
%%==================================%%

\abstract{We study the  monotone  nonlinear energy stability of \textit{magnetohydrodynamics  plane  shear flows, Couette and Hartmann flows}.  We prove that the least stabilizing perturbations, in the energy norm, are the two-dimensional spanwise perturbations and give some criti\-cal Reynolds numbers  Re$_E$ for some selected Prandtl and Hartmann numbers. This result solves a conjecture given in a recent paper by Falsaperla et al. \cite{FMP.2022} and implies a Squire theorem for nonlinear energy: the less stabilizing perturbations in the \textit{energy norm} are the two-dimensional spanwise perturbations. Moreover, for Reynolds numbers less than Re$_E $ there can be no transient energy growth.}

\keywords{Magnetic Couette flow, Hartmann flow, Nonlinear monotone energy stability. }

%%\pacs[JEL Classification]{D8, H51}

\pacs[MSC Classification]{76E05, 76E25}

\maketitle

\textit{Dedication.\\
This work is dedicated to Prof. Salvatore Rionero, my dearest teacher and mentor. His memory will always remain in my heart forever.}

\section{Introduction}\label{sec1}

It is well known that the study of the stability of laminar flows in magnetohydrodynamics is important for the numerous applications to different fields: geophysics, astrophysics, industry, biology, in metallurgy, in biofilms, and medicine, see \cite{ Alexakis_etal.2003} -  \cite{Davidson.2001}, and the references therein.

Many stability problems in magnetohydrodynamics even in the presence of temperature have been studied and some notable results have been obtained by Rionero \cite{Rionero.1967} - \cite{Capone.Rionero.2016}, also in porous media.  In particular,  in the work \cite{Rionero.1968}, Rionero proves, in the magnetohydrodynamics case, the fundamental existence theorem of the maximum of a functional ratio connected to the Reynolds-Orr energy equation.

In a recent paper Falsaperla et al. \cite{FMP.2022},  studied the monotone nonlinear energy stability of Couette and Hartmann motions with respect to three-dimensional perturbations in magnetohydrodynamics. They found that the streamwise perturbations are stabilizing for any Reynolds number. This is in  contradiction with the results of  Alexakis et al. \cite{Alexakis_etal.2003}.  In order to solve
this contradiction, Falsaperla et al. \cite{FMP.2022} made a conjecture: the maximum of the functional ratio that comes from the Reynolds-Orr energy equation is obtained in a subspace of the space of kinematically admissible perturbations, the space of physically admissible perturbations competing for the maximum, the two-dimensional spanwise perturbations.

The main purpose of this paper is to prove that this conjecture is true: the  maximum of the functional ratio that comes from the Reynolds-Orr energy
equation, and consequently the critical nonlinear Reynolds number for monotone energy stability, is obtained on two-dimensional perturbations, the spanwise perturbations. To obtain this result, we write the Reynolds-Orr energy equation and, as Lorentz \cite{Lorentz.1907}, ( see also \cite{Lamb.1924}), has observed in the fluid-dynamics case, we remark that a \textit{scale invariance property} holds for the terms of the energy equation. Then, we compare two functional ratios and study the maximum obtained with the Euler-Lagrange equations.

The plan of the paper is the following. In Section 2 we introduce the basic motions and the perturbation equations.
In Section 3 we study the nonlinear energy stability with respect to three-dimensional perturbations and find that the critical Reynolds numbers for monotone energy stability are obtained on the spanwise two-dimensional perturbations.

In Section 4 we report some graphs of the critical  Reynolds numbers obtained with the Chebyshev collocation method for fixed Prandtl and Hartmann numbers. Finally, in section 5, we draw a conclusion.

\section{Basic motions and perturbation equations}\label{sec2}

Consider a layer $\mathcal D= \mathbb R^2 \times [-1,1]$ filled with an electrically conducting fluid, \cite{Davidson.2001}. 
We can write the magnetohydrodynamics system for stationary flows in the non-dimensional form \cite{Davidson.2001, Takashima.1996}, and \citep[formula (14)]{Falsaperla.Mulone.Perrone.2021a}: 
\begin{eqnarray}\label{basicmotion-nd-a}
	\left\{ \begin{array}{l}\label{main}
		{\bf v}\!\cdot\!\nabla{\bf v} = \Ha^2 \Re^{-1} \Rm \,{\bf \hat B}\!\cdot\!\nabla{\bf \hat B} - \nabla \Pi + \Re^{-1}\Delta{\bf v}\\
		\nabla\!\cdot\!{\bf v}=0\\
		{\bf v}\!\cdot\!\nabla {\bf \hat B} -{\bf \hat B}\!\cdot\!\nabla{\bf v} = \Rm^{-1}\,\Delta{\bf \hat B}\\
		\nabla\!\cdot\!{\bf \hat B}=0,
	\end{array}  \right.
\end{eqnarray}
where $(x,y,z) \in \mathcal D$ and ${\bf v}$, ${\bf \hat B}$  are the unknown fields, respectively the velocity of the fluid, the magnetic induction field, and $\Pi$ is the effective pressure (including the magnetic pressure). They are regular fields (at least $C^2(\mathcal D)$). The other symbols in (\ref{basicmotion-nd-a}) are the positive non-dimensional parameters 
\begin{itemize}
	\item $\Re= V_0d/\nu$,  the Reynolds number,
	\item $\Rm= V_0d/\eta$,  the magnetic Reynolds number,
	\item $\Pm = \dfrac{\nu}{\eta}= \dfrac{\Rm}{\Re}$, the magnetic Prandtl number,
	\item $\Ha= \dfrac{B_0 d}{\sqrt{\rho \nu \mu \eta}}$, the Hartman number.
\end{itemize}
$V_0$ and $B_0$ are a reference velocity (generally the maximum velocity is considered) and a reference magnetic field.  $d$, $\nu$, $\eta$, $\rho$, $\mu$  are the half width of the layer,  the viscosity, the electric resistivity, the density and the magnetic permeability, respectively; they are positive numbers. $\nabla$ is the gradient operator and $\Delta$ is the three-dimensional Laplacian.

Following \cite{Falsaperla.Giacobbe.Mulone.2020} we restrict our analysis to $z$-dependent laminar solutions of the form (we call them \textit{mean or basic} solutions)

\[ \label{bs-1}{\bf{v}}(z)=(U(z), 0,0), \quad {\bf \hat B}(z)=(\bar B(z), 0, \Rm^{-1})
\]
and we choose boundary conditions  for plane \textit{Couette and Hartmann flows} which correspond to  rigid conditions for the kinetic field and non-conducting boundaries, (cf.  \cite{Alexakis_etal.2003}). 
We also assume that there is no forcing pressure in the channel.

We recall the following Theorems (see \cite{Alexakis_etal.2003}, \cite{Takashima.1996}, \cite{Takashima.1998}, \cite{Falsaperla.Giacobbe.Mulone.2020}):

\begin{thm}
	The basic solution of system \eqref{basicmotion-nd-a} satisfying the boundary conditions 
	\[ \label{cc-rnc}
	U(-1)= -1, \qquad U(1) = 1, \qquad \bar B(-1) = \bar B(1) = 0
	\]	is the magnetic \textit{Couette} flow
	\[\label{Couette}
	U(z) = \dfrac{\sinh (\Ha\,z) }{\sinh\left(\Ha\right)}, \quad \quad \bar B(z)= \dfrac{\cosh\left(\Ha\right)- \cosh (\Ha\, z)}{\Ha\sinh\left(\Ha\right)}
	\]
\end{thm}

\begin{thm}
	The basic solution of system \eqref{basicmotion-nd-a} satisfying the boundary conditions 
	\[
	U(-1) =  U(1) =0, \qquad \bar B(-1) = \bar B(1) = 0
	\]
	is the \textit{Hartmann} flow
	\[\label{Hartmann}
	U(z) = \dfrac{\cosh (\Ha) - \cosh (\Ha\, z) }{\cosh (\Ha)-1},\quad \bar B(z) = \dfrac{ \sinh(\Ha\, z)- z \sinh (\Ha) }{\Ha(\cosh (\Ha)-1)}.
	\]
\end{thm}
We note that, with the given values of ${\bf{v}}(z)$ and ${\bf \hat B}(z)$, the pressure $\Pi$ can be obtained by solving \eqref{main}$_1$ with respect to $\Pi$. 

\vskip .3cm
We want to investigate the nonlinear stability of these basic solutions. To this end, we consider a  regular ($C^2(\mathcal D \times [0, +\infty)$) disturbance of the stationary solution
\[
{\bf v} + {\bf u}= (U(z),0,0) +(u, v, w),\quad {\bf \hat B + h} = (\bar B(z), 0, \Rm^{-1}) + (h, k, \ell), \quad \Pi + \bar\pi,
\]
with $(u, v, w)$, $(h, k, \ell)$ and $\bar\pi$ depending on the variables $x,y,z$, and $t$.

Denoting with  
\be\label{A} A = \Ha^2 \Re^{-1} \Rm= \Ha^2 \Pm,
\ee
the equations which govern the evolution of the \textit{difference fields} ${\bf u}, {\bf h}, \bar \pi$ (often such difference fields are improperly  called perturbations or disturbances) are:
\begin{equation}\label{perturb2}
	\begin{cases}
		{\bf u}_t+ U(z) {\bf u}_x + w U'(z) {\bf i} + {\bf u}\cdot \nabla {\bf u} =   A[\bar B(z) {\bf h}_x+  \dfrac{{\bf h}_z}{\Rm}+\\[10pt]+ \ell \bar B'(z) {\bf i}+ {\bf h}\cdot \nabla {\bf h}]-\nabla \bar \pi + \dfrac{\Delta{\bf u}}{\Re} \\[10pt]
		{\bf h}_t+w \bar B'(z) {\bf i}+U(z) {\bf h}_x + {\bf u}\cdot \nabla {\bf h}-\bar B(z) {\bf u}_x- \dfrac{{\bf u}_z}{\Rm}- \ell U'(z) {\bf i}+\\[10pt]- {\bf h}\cdot \nabla {\bf u}= \dfrac{\Delta{\bf h}}{\Rm} \\[10pt]
		\nabla\!\cdot\!{\bf u}=0, \quad \nabla\!\cdot\!{\bf h}=0\, ,
	\end{cases}
\end{equation}
where the suffixes $t$, $x$ and $z$ denote derivatives with respect to the corresponding variables, the superscript denotes first derivative with respect to $z$.

We assume that the perturbations are periodic in the variables $x$ and $y$,  denote with $\Omega=[0,\dfrac{2\pi}{a}]\times[0,\dfrac{2\pi}{b}]\times[-1,1]$ a periodicity cell \cite{Falsaperla.Giacobbe.Mulone.2020}, and denote with $L_2(\Omega)$ the space of real square-integrable functions in $\Omega$.
We indicate with the symbols $(\cdot, \cdot )$ and $\Vert \cdot \Vert$ the usual scalar product and the norm in  the space of square-summable functions in $\Omega$, $L_2(\Omega)$.

The most common boundary conditions for ${\bf u}, {\bf h}$  on the planes $z=\pm1$ are  (see Chandrasekhar \cite{Chandrasekhar.1961})
\begin{enumerate}
	\item rigid (\textit{r}), $u = v = w = 0$
	\item stress-free (\textit{sf}), $u_z = v_z = w = 0$
	\item non-conducting (\textit{n}), $h = k = \ell = 0 $
	\item conducting (\textit{c}), $h_z = k_z = \ell = 0 .$
\end{enumerate}

Here we consider only the rigid and non-conducting case. Other boundary conditions will be consider in future papers. 

We recall the (usual) definitions of streamwise and spanwise perturbations:
\begin{defn}
	{The perturbations  \textit{streamwise} (or longitudinal) are perturbations ${\bf u}, {\bf h}, \bar\pi$  which do not depend on $x$. 
}\end{defn}

\begin{defn}
	{The perturbations \textit{ spanwise} (or transverse) are  perturbations ${\bf u}, {\bf h}, \bar\pi$   which do not depend on $y$.} The two-dimensional spanwise perturbations are the spanwise perturbations with $v=k=0$.
\end{defn}

\section{Nonlinear energy stability}\label{sec4}
First we recall the main nonlinear energy stability definitions.
\vskip .3cm
\begin{defn}
	We define the \textit{energy} (see \cite{Falsaperla.Giacobbe.Mulone.2020}) of a disturbance ${\bf u}, {\bf b}$, 
	$$E(t) = \dfrac{1}{2} (\Vert {\bf u}\Vert^2 +   A\Vert {\bf h}\Vert^2), $$
	with the coupling parameter $A$ given by \eqref{A}. 
\end{defn}

\begin{defn}
	A basic motion ${\bf{v}}(z)=(U(z), 0,0), \quad {\bf \hat B}(z)=(\bar B(z), 0, \Rm^{-1})$ is\textit{ monotone stable in the energy norm} $E$ of a disturbance, and ${\rm Re}_E$ is the  critical Reynolds number, if the time orbital derivative of the energy, $\dot E$, is always less than zero,
	\be \dot E <0 ,\ee
	when $ {\rm Re}< {\rm Re}_E $. In particular the stability is monotone and exponential decreasing if there is a positive number $\alpha$ such that  $E(t)\le E(0) \exp\{-\alpha t\}$ for any $t\ge 0$ and $ {\rm Re}< {\rm Re}_E $.
\end{defn}

\begin{defn}
	A basic motion  ${\bf{v}}(z)=(U(z), 0,0), \quad {\bf \hat B}(z)=(\bar B(z), 0, \Rm^{-1})$  to the Navier-Stokes magnetohydrodynamics equations
	is \textit{globally stable} to perturbations if the perturbation energy $E$ satisfies
	
	\be \label{Global} \lim_{t\to +\infty} \dfrac{E(t)}{E(0)} =0, \quad \forall \,  E(0)>0.\ee
\end{defn}

Now we study (and recall some results in \cite{Falsaperla.Giacobbe.Mulone.2020})  the nonlinear stability of the shear flows by using the Lyapunov second method with the classical energy (see \cite{Falsaperla.Giacobbe.Mulone.2020})
$$E(t) = \dfrac{1}{2} (\Vert {\bf u}\Vert^2 +   A\Vert {\bf h}\Vert^2) . $$

Taking the orbital derivative of $E(t)$ and considering equations \eqref{perturb2}, the periodicity, the boundary conditions and the solenoidality  of ${\bf u}$ and ${\bf h}$, we obtain the Reynolds-Orr \cite{Reynolds.1883}, \cite{Orr.1907} equation (see \cite{Falsaperla.Giacobbe.Mulone.2020})
\begin{align}\nonumber
\dot E = -(wU',u) +   A\left[(\ell \bar B',u) -(w \bar B',h) + (\ell U',h) \right]+\\ - \Re^{-1} \Vert \nabla {\bf u} \Vert^2
-   A\,\Rm^{-1} \Vert \nabla {\bf h} \Vert^2.
\end{align}

As in the case of fluid-dynamics (see Lorentz  \cite{Lorentz.1907},  Lamb \cite{Lamb.1924}, p. 640) we note that ``the relative magnitude of the two terms on the right-hand side  is unaffected if we reverse the signs of $u, v, w$, and of $h$, $k$ and $\ell$ or if we multiply them by any constant factor. The stability of a given state of mean motion should not therefore depend on the \textit{scale} of the disturbance" (the constant factor must be the same for ${\bf u}$ and ${\bf h}$). Therefore, in the study of the following maximum problems we will always assume that this \textit{scale invariance} property holds.

\subsection{Nonlinear stability with respect to three-dimensional perturbations}\label{sec3}

Applying classical methods, see \cite{Rionero.1968, Joseph.1976, Straughan.2004}, 
we define 
\be\label{I}
I=-(U' w,u)+   A\left[(\ell \bar B',u) -(w \bar B',h) + (\ell U',h)\right] ,
\ee 
and assume that the perturbations satisfy the conditions ${\bf u}=0$ and ${\bf b}=0$ on the boundaries, are divergence-free, periodic in $x$ and $y$,  and they satisfy the condition $ \Vert \nabla {\bf u} \Vert+ \Vert \nabla {\bf h} \Vert>0$, and the scale invariance property. We can write the energy equation in this way
\be \label{en1}
\dot E = I- \Re^{-1} \Vert \nabla {\bf u} \Vert^2 -   A\Rm^{-1} \Vert \nabla {\bf h} \Vert^2 .
\ee
Introducing the space ${\cal S}$ of the kinematically admissible perturbations ${\bf u}$ and ${\bf h}$ periodic in $x$ and $y$, 
\be \begin{array}{l}\label{spaceS}
	{\cal S}= \{{\bf u}, {\bf h}\in W^{2,1} (\Omega), \; {\bf u}={\bf h}=0 \hbox{ when } z=\pm 1, \nabla \cdot {\bf u}=\nabla \cdot {\bf h}=0,  \; \\ \qquad \Vert \nabla {\bf u} \Vert+ \Vert \nabla {\bf h} \Vert>0\},
\end{array}
\ee
where $W^{2,1} (\Omega)$ is the Sobolev space defined as the subspace of the space of vector fields with their components  $f_i$  ($i=1,2,3$) in $L_{2}(\Omega)$ such that $f_i$ and its weak derivatives up to the first order  have a finite $L_2$-norm.

In order to solve the conjecture made in \cite{FMP.2022}, we use the method given in \cite{Mulone.2023}.

Firstly, we observe that in the case $I \le 0$ we have $\dot E<0$, and the perturbations are monotonically stable. 

If instead \textit{$I$ is greater than zero}, then  for any perturbation in 	${\cal S}$ that satisfy the scale invariance property,  we may write \eqref{en1} in the following way

\be \label{en2-2}
\dot E= \left[\dfrac{I}{ \Vert \nabla {\bf u} \Vert^2 +   \Ha^2 \Vert \nabla {\bf h} \Vert^2} -\Re^{-1}\right] (\Vert \nabla {\bf u} \Vert^2 +   \Ha^2 \Vert \nabla {\bf h} \Vert^2).
\ee

In the case \textit{$I$ greater than zero}, for any perturbation ${\bf u}, {\bf h}$ in $\cal S$, we have

\be \label{inq1}
\dfrac{I}{ \Vert \nabla {\bf u} \Vert^2 +   \Ha^2 \Vert \nabla {\bf h} \Vert^2}
\le \dfrac{I}{ \Vert \nabla {u} \Vert^2 +  \Vert \nabla {w} \Vert^2  +\Ha^2  [\Vert \nabla {h} \Vert^2 + \Vert \nabla {\ell} \Vert^2]}.
\ee

Defining
$${\cal D}_1= \Vert \nabla {u} \Vert^2 +  \Vert \nabla {w} \Vert^2  +\Ha^2  [\Vert \nabla {h} \Vert^2 + \Vert \nabla {\ell} \Vert^2], $$
and 
$${\cal D}= \Vert \nabla {\bf u} \Vert^2 +\Ha^2  [\Vert \nabla {\bf h} \Vert^2 ], $$
we now prove that 
\be \label{egua-max}\max_{\cal S} \dfrac{I}{\cal D} = \max_{{\cal S}_0} \dfrac{I}{{\cal D}_1},\ee where ${\cal S}_0$ is the subspace of ${\cal S}$ of the two-dimensional spanwise perturbations.

To see this, we choose any element $({\bf u}, {\bf h})$ in ${\cal S}$, we have

\be
\dfrac{I}{ \cal D}\le \dfrac{I}{{\cal D}_1}\le \max_{\cal S} \dfrac{I}{{\cal D}_1}.
\ee
From this inequality it follows that $ \max_{\cal S} \dfrac{I}{{\cal D}_1}$ is an upper bound of the set of elements $\dfrac{I}{ \cal D}$ when $(\bf u, h)$ vary in $\cal S$. Therefore, $ \max_{\cal S} \dfrac{I}{\cal D}$ is the \textit{least} upper bound, and
$$ \max_{\cal S} \dfrac{I}{\cal D} \le  \max_{\cal S} \dfrac{I}{{\cal D}_1}.$$
Finally, in the next subsection we shall prove that
\be \label{finale}\max_{\cal S} \dfrac{I}{{\cal D}_1} = \max_{{\cal S}_0} \dfrac{I}{{\cal D}_1}.\ee
 This implies \eqref{egua-max}, because ${\cal S}_0$ is a subspace of ${\cal S}$.

Assuming ${\cal D}_1>0$ and observing that the Poincaré's inequality holds, we have that the ratio  $\dfrac{I}{{\cal D}_1}$ is bounded from above in ${\cal S}$.  A theorem due to Rionero \cite{Rionero.1968} (see also Galdi and Rionero \cite{Galdi.Rionero.1985}) proves that the functional ratio

$${\cal F}= \dfrac{I}{{\cal D}_1}=\dfrac{I}{\Vert \nabla {u} \Vert^2 +  \Vert \nabla {w} \Vert^2  +\Ha^2  [\Vert \nabla {h} \Vert^2 + \Vert \nabla {\ell} \Vert^2]}  $$ admits a maximum in ${\cal S}$.

Denoting this maximum with 
\be\label{max}
{\rm Re}_E^{-1} = m = \max_{\cal S} \dfrac{-(U' w,u)+   A\left[(\ell \bar B',u) -(w \bar B',h) + (\ell U',h)\right] }{\Vert \nabla {u} \Vert^2 +  \Vert \nabla {w} \Vert^2  +\Ha^2  [\Vert \nabla {h} \Vert^2 + \Vert \nabla {\ell} \Vert^2]},
\ee
from \eqref{en2-2},  \eqref{inq1} and \eqref{max}, we have the inequality
\be\label{en-id1}
\dot E 
\le ({\rm Re}_E^{-1} - \Re^{-1})  [ \Vert \nabla {\bf u} \Vert^2 +  \Ha^2 \Vert \nabla {\bf h} \Vert^2 ].
\ee

From this inequality and the Poincaré's inequalities $$\dfrac{\pi^2}{4} \Vert {u}\Vert^2 \le \Vert \nabla {u} \Vert^2, \quad \dfrac{\pi^2}{4} \Vert {v}\Vert^2 \le \Vert \nabla {v} \Vert^2, \quad \dfrac{\pi^2}{4} \Vert {w}\Vert^2 \le \Vert \nabla {w} \Vert^2,$$ $$ \dfrac{\pi^2}{4} \Vert {h}\Vert^2 \le \Vert \nabla {h} \Vert^2, \quad \dfrac{\pi^2}{4} \Vert {k}\Vert^2 \le \Vert \nabla {k} \Vert^2, \quad \dfrac{\pi^2}{4} \Vert {\ell}\Vert^2 \le \Vert \nabla {\ell} \Vert^2, $$
it follows that  condition
$$\Re < {\rm Re}_E$$ implies nonlinear monotone energy stability of magnetic Couette and  Hartmann  motions:

\begin{thm}
	Assuming that the Reynolds number satisfies condition 
	$$\Re < {\rm Re}_E, $$  the basic magnetic Couette and Hartmann motions are monotone asympotically stable in the energy norm $E$ according to the inequality
	$$E(t) \le E(0) e^{\frac{\pi^2}{2} c_0(\Re-{\rm Re}_E)t},$$
	with a positive constant $c_0$ depending on $\Ha$ and $\Pm$.
\end{thm}

\subsection{Nonlinear critical Reynolds number}

We prove here that the nonlinear critical Reynolds number is obtained on two-dimensional spanwise disturbances (the \textit{Orr perturbations} in fluid dynamics).

In order to compute the critical Reynolds number for the monotone nonlinear energy stability, we have to compute ${\rm Re}_E$ or $m= 1/{\rm Re}_E$. For this purpose we must write the Euler-Lagrange equations of the functional ${\cal F}$.

The Euler-Lagrange equations of the maximum problem \eqref{max} are (see \cite{Falsaperla.Giacobbe.Mulone.2020}, \cite{Joseph.1976})
\begin{eqnarray}\label{EL}
	\left\{ \begin{array}{l}
		[-U'w{\bf i}- U' u {\bf k}+   A \bar B' \ell  {\bf i} -  A \bar B' h {\bf k}]+2 m (\Delta {u} {\bf i} + \Delta {w} {\bf k}) = \nabla \lambda_1\\[5pt]
		A [\bar B' u{\bf k}- w \bar B' {\bf i}+ U' \ell {\bf i}+ U' h {\bf k}] + 2 m \Ha^2 (\Delta {h}{\bf i} +\Delta {\ell}{\bf k} ) =  \nabla \lambda_2 ,
	\end{array}  \right.
\end{eqnarray}
where $\lambda_1$ and $\lambda_2$ are Lagrange multipliers which depend on $x,y,z$.

We can write the Euler-Lagrange equations in components
\begin{eqnarray}\label{EL-comp0}
	\left\{ \begin{array}{l}
		-U'w +   A \bar B' \ell +2 m \Delta {u}  = \dfrac{\partial \lambda_1}{\partial x}\\[10pt]
		\qquad \qquad \qquad \qquad \qquad 0= \dfrac{\partial \lambda_1}{\partial y}\\[10pt]
		- U' u  -  A \bar B' h +2 m  \Delta {w}  = \dfrac{\partial \lambda_1}{\partial z}\\[10pt]
		A [- w \bar B' + U' \ell] + 2 m \Ha^2 \Delta {h}  =  \dfrac{\partial \lambda_2}{\partial x} \\[10pt]
		\qquad \qquad \qquad \qquad \qquad  0= \dfrac{\partial \lambda_2}{\partial y}\\[10pt]
		A [\bar B' u+ U' h ] + 2 m \Ha^2 \Delta {\ell}  =  \dfrac{\partial \lambda_2}{\partial z}  \\[5pt]
		u_x+v_y+w_z=0, \; h_x+k_y+\ell_z=0,
	\end{array}  \right.
\end{eqnarray}
therefore the two multipliers $\lambda_1$ and $\lambda_2$ do not depend on $y$.

If $\lambda_1=0$ and $\lambda_2=0$, we take into account the conditions of solenoidality $u_x+v_y+w_z=0$, $h_x+k_y+\ell_z=0$ and the boundary conditions $w=w_z=\ell=\ell_z=v=k=0$  (the boundary conditions for $w_z$ and for $\ell_z$ are obtained from the solenoidality of ${\bf u}$  and ${\bf b}$  and from the boundary conditions for $u$, $v$, $h$ and $k$.) Then, we take the successive derivatives with respect to $z$ of each equation of \eqref{EL-comp0}. It is not difficult to prove that $u, w, h, \ell$ and all their derivatives with respect to $z$ are zero on the boundary, therefore $u, w$ and $h, \ell$ are identically zero. This implies that ${\bf u}= {\bf h}=0$ in $\Omega$ that has been excluded.

If  $\lambda_1$ and $\lambda_2$ are non-zero functions dependent on $x$ and $z$, taking into account that $m$ is the maximum in \eqref{max}, 
it is not difficult to prove (see \cite{Mulone.2023}) that $v=0$ and $k=0$, and now the solenoidality  conditions are $u_x+w_z=h_x+\ell_z=0$. We derive \eqref{EL-comp0}$_1$ with respect to $z$ and \eqref{EL-comp0}$_3$ with respect to $x$, we subtract and take into account the conditions of solenoidality \eqref{EL-comp0}$_7$. Likewise,  we derive \eqref{EL-comp0}$_4$ with respect to $z$ and \eqref{EL-comp0}$_6$ with respect to $x$, we subtract and take into account the conditions of solenoidality \eqref{EL-comp0}$_7$. We obtain
\be\label{elwz-0}
\begin{cases}
	-U' w_{z}- U''w  + A \bar B''\ell + A \bar B' \ell_z+2 m \Delta u_z+U' u_x+ A \bar B' h_x-2m \Delta w_x=0\\
	AU'  \ell_{z}+ A U''\ell  - A \bar B''w -  A \bar B'w_z- A \bar B' u_x- A U' h_x+ 2m \Ha^2 \Delta h_z+\\ -2 m \Ha^2 \Delta \ell_x=0.
\end{cases}
\ee
By differentiating each of the equations with respect to $x$ and applying the conditions of solenoidality, we finally have:
\be\label{el-finale}
\begin{cases}
	2U' w_{xz}+ U''w_x  - A \bar B''\ell_x + 2 m \Delta  (w_{xx}+ w_{zz})=0\\
	-2A U' \ell_{xz}- A U''\ell_x  + A \bar B''w_x + 2 m \Ha^2\Delta (\ell_{xx}+ \ell_{zz}) =0,
\end{cases}
\ee
with the boundary conditions 
\be\label{elwz-bc0}
w=w_z=\ell=\ell_z=0, 
\ee
on $z=\pm 1$.

We observe that, as it is easy to check, the maximum  of \eqref{max} is obtained when $u_y=0$, $w_y=0$, $h_y=0$ and $\ell_y=0$. Therefore, $u, w, h, \ell$ depend only on $x$ and $z$, and  in \eqref{el-finale} we have $\Delta= \dfrac{\partial^2}{\partial x^2}+ \dfrac{\partial^2}{\partial z^2}$. 

Since system \eqref{el-finale} is linear, we seek solution of the form
(see \cite{Chandrasekhar.1961, Joseph.1976, Drazin.Reid.2004, Straughan.2004}):
\begin{equation} \label{pf0} F(x,y,z) = f(z) e^{ia x} \, ,\end{equation}
with $F =  w, \ell$ in the domain $\Omega$. We have the system
\be\label{el-finale-a}
\begin{cases}
	2iaU' Dw+ ia U''w  - ia A \bar B''\ell + 2 m (D^2-a^2)^2 w=0\\
	-2ia A U' D\ell- ia A U''\ell  + ia A \bar B''w + 2 m \Ha^2 (D^2-a^2)^2 \ell =0,
\end{cases}
\ee
with $D= \dfrac{d}{d z}$.

System \eqref{el-finale-a} is the \textit{Orr system} for shear flows \textit{in magnetohydrodynamics}. 
This ordinary linear differential system with coefficients that depend on $z$ is an eigenvalue problem for $m$ (or ${\rm Re}_E$). 

The critical Reynolds numbers we obtain from this system correspond exactly to the critical Reynolds numbers obtained by Orr \cite{Orr.1907} (see also recent results of  Falsaperla et al. \cite{FMP.2022}) in fluid dynamics (i.e. the critical Reynolds numbers are reached for the \textit{two-dimensional spanwise} perturbations). The Orr's system in fluid dynamics is formally obtained from  \eqref{el-finale-a} by setting therein $\Ha=0$.

This result proves \textit{the relation \eqref{finale} holds}, and \eqref{egua-max} is shown.

A consequence of this result is that a Squire theorem, \cite{Squire.1933}, holds for nonlinear energy stability in magnetohydrodynamics: the less stabilizing perturbations in the energy norm are the two-dimensional spanwise  perturbations. 
We observe that in \textit{the linear case}  Takashima, \cite{Takashima.1996}, and {\cite{Takashima.1998}, p. 109,   writes
	``It is evident from Eqs. (2.30)-(2.32) and the boundary conditions (2.33)-(2.35) \textit{that Squire's theorem is valid,} and therefore we shall hereafter consider only two-dimensional disturbances in the
	$z-x$ plane (i.e., $\beta = 0$).}"

\section{Some numerical results}

We show here some numerical results. These results are obtained solving system (\ref{el-finale-a}) with boundary condition (\ref{elwz-bc0}). Eigenvalue problem (\ref{el-finale-a})-(\ref{elwz-bc0}) has been solved in \cite{FMP.2022} with a Chebyshev collocation method, using between 50 and 70 base polynomials. For completeness, we report here their result for spanwise perturbations. We fix $\Pm=0.1$ and $\Ha=0.1, 1, 10, 50$.

\begin{figure}[H]
	\begin{center}
		\includegraphics[width=5.7cm]{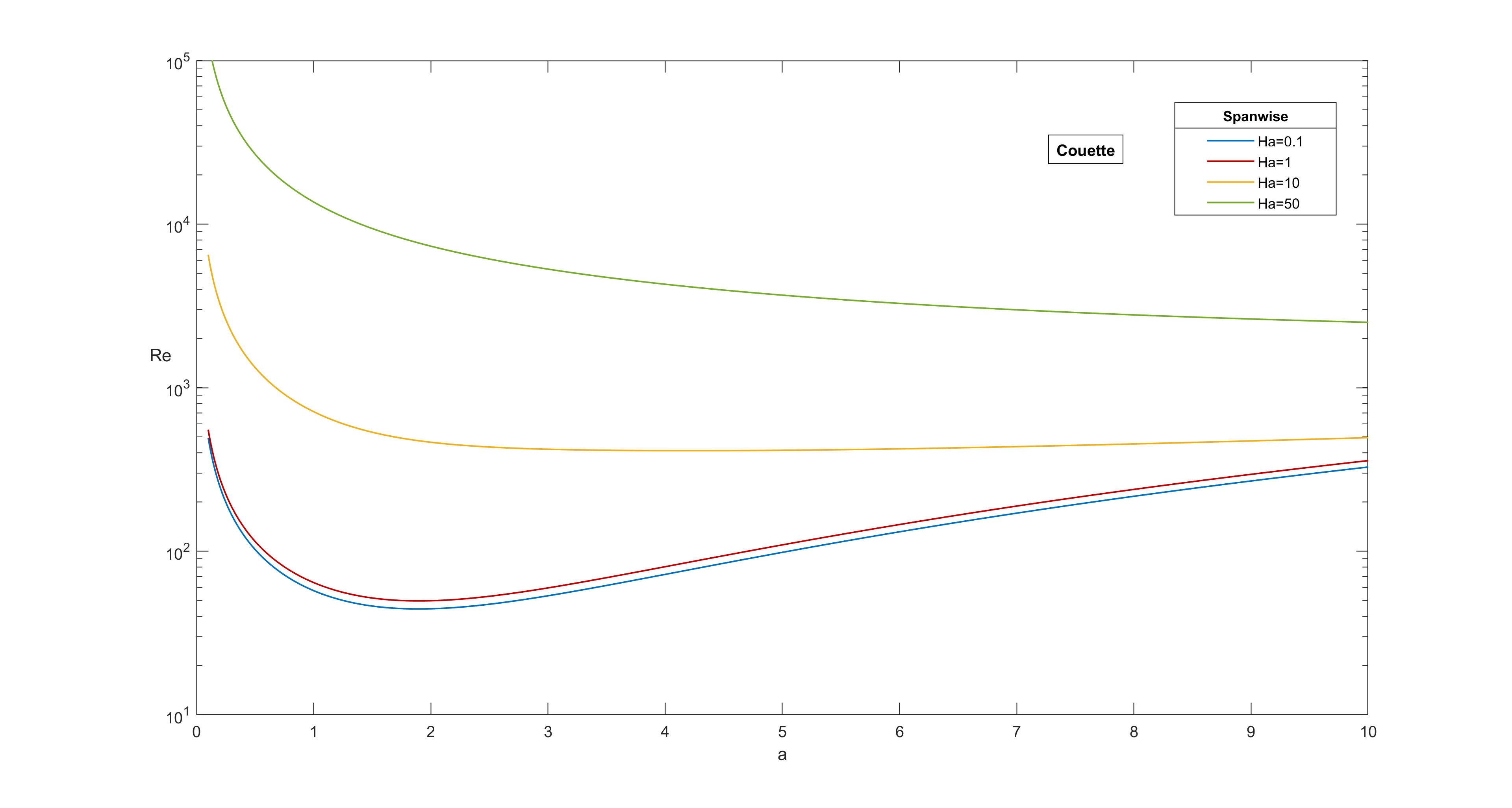}  \includegraphics[width=5.7cm]{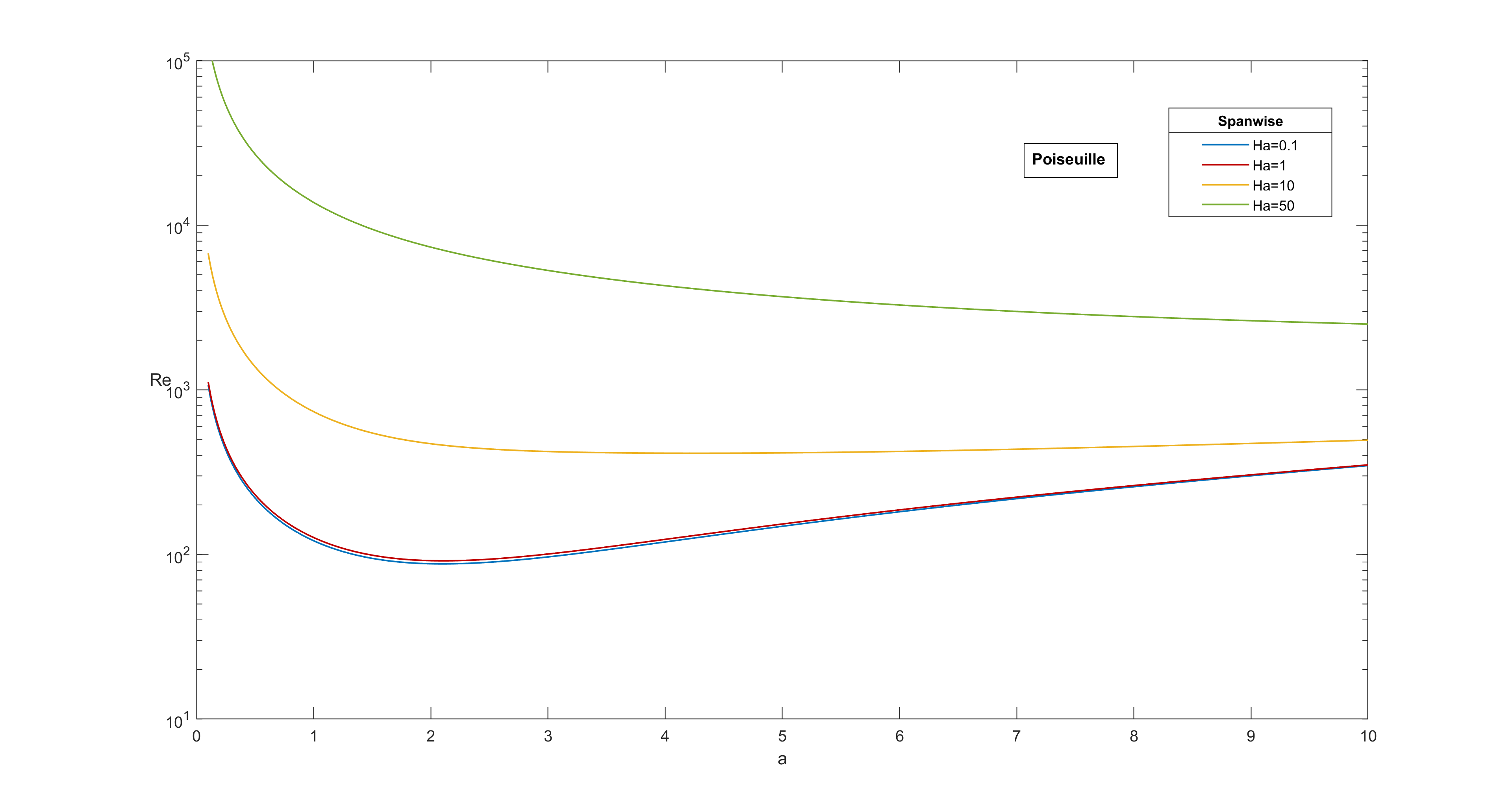}
	\end{center}
	\caption{Orr-Reynolds critical number ${\rm Re}$ for magnetic Couette flow (left panel) and Hartmann (magnetic Poiseuille) flow (right panel) as a function of the wave number $a$ and magnetic Prandtl number ${\rm Pm}=0.1$. }
	\label{Orr-ab}
\end{figure}	

In Fig.~\ref{Orr-ab} we fix $\Pm=0.1$ and $\Ha=0.1, 1, 10, 50$ and we obtain the Reynolds number $\Re$ as a function wave number $a$. For each fixed value of $\Ha$ the critical Reynolds value, $\Re_E$ is found taking the minimum of $\Re$ with respect to the parameter $a$.

\section{Conclusion}

In this paper we study the monotone nonlinear energy stability of \textit{magnetohydrodynamics plane Couette and Hartmann} shear flows with rigid and non-conducting boundaries.

We solve the conjecture given in \cite{FMP.2022} proving that the nonlinear critical Reynolds number is obtained on spanwise perturbations. To prove this we compare two functional ratios and take into account that the second member of the energy equation has a scale invariance property  with respect to the fields ${\bf u}$ and ${\bf h}$. We therefore solve the Euler-Lagrange equations and prove that the maximum is obtained on the functions which have $v=0$, $k=0$ and do not depend on $y$. 

This results implies a Squire theorem for nonlinear stability. Moreover, for Reynolds numbers less than Re$_E $ there can be no transient energy growth.

\bmhead{Acknowledgments}

I thank Dr. Carla Perrone for making the graphs in section 4.

The research that led to the present paper was partially supported by the following Grants: 2017YBKNCE of national project PRIN of Italian Ministry for University and Research, by  a grant: PTR-DMI-53722122113 ``Analisi qualitativa per sistemi dinamici finito e infinito dimensionali con applicazioni a biomatematica, meccanica dei continui e termodinamica estesa classica e quantistica" of University of Catania. I also thank the group GNFM of INdAM for financial support. 

\textbf{Declarations}

Conflicts of interest: The author states that there is no conflict of
interest.

\end{document}